\documentclass[iop]{emulateapj}
\usepackage{latexsym,amssymb}
\usepackage{natbib}
\usepackage{hyperref}

\newcommand{\Msun}{\ensuremath{{\rm M}_\odot}}
\newcommand{\Lsun}{\ensuremath{{\rm L}_\odot}}
\newcommand{\Rsun}{\ensuremath{{\rm R}_\odot}}
\newcommand\lsim{\mathrel{\rlap{\lower4pt\hbox{\hskip1pt$\sim$}}
        \raise1pt\hbox{$<$}}}
\newcommand\gsim{\mathrel{\rlap{\lower4pt\hbox{\hskip1pt$\sim$}}
        \raise1pt\hbox{$>$}}}
\def\Bnabla{\mbox{\boldmath$\nabla$}}

\shorttitle{Stellar Structure and Modified Gravity}
\shortauthors{Chang \& Hui}

\begin{document}

\title{Stellar Structure and Tests of Modified Gravity} 


\author{Philip Chang\altaffilmark{1} \& Lam Hui\altaffilmark{2}}

\altaffiltext{1}{Canadian Institute for Theoretical Astrophysics, 60 St George St, Toronto, ON M5S 3H8, Canada, email: pchang@cita.utoronto.ca}
\altaffiltext{2}{Department of Physics and Institute for Strings, Cosmology and Astroparticle Physics, Columbia University, New York, NY 10027, email: lhui@astro.columbia.edu}

\begin{abstract} 
Theories that attempt to explain cosmic acceleration by
modifying gravity typically introduces a long-range scalar force 
that needs to be screened on small scales.
One common screening mechanism is the chameleon, where 
the scalar force is screened in environments with a 
sufficiently deep gravitational potential, but
acts unimpeded in regions with a shallow gravitational potential.
This leads to a variation in the overall gravitational G
with environment. We show such a variation
can occur within a star itself, significantly affecting
its evolution and structure, provided that the host galaxy is unscreened.
The effect is most pronounced for
red giants, which would be smaller by a
factor of tens of percent and thus hotter by 100's of K, 
depending on the parameters of the underlying scalar-tensor theory.
Careful measurements of these stars in suitable environments
(nearby dwarf galaxies not associated with groups or clusters) 
would provide constraints on the chameleon mechanism that 
are four orders of magnitude better than current large scale
structure limits, and two orders of magnitude better
than present solar system tests.  
\end{abstract}

\begin{keywords}
{
stars: evolution -- cosmological parameters -- cosmology: theory
}
\end{keywords}

\section{Introduction}\label{sec:intro}

The discovery of cosmic acceleration a decade ago has spurred a number
of attempts to modify general relativity (GR) on large
scales. Such attempts generally take the form of a scalar-tensor theory,
at least in appropriate limits (e.g. scales small compared to Hubble).
The fact that an extra (scalar) force is inevitably introduced is a consequence
of a theorem due to \cite{Weinberg1965} and \cite{Deser1970}, 
which states that a Lorentz-invariant theory of a massless spin-two particle
must be GR at low energies, or long distances. 
In other words, to modify the long range interaction between
masses, one has no other option but to introduce extra degrees of
freedom, such as a scalar field.
The force mediated by this scalar must be suppressed or screened
in an environment such as the solar system, to satisfy stringent
experimental constraints. One common screening mechanism is
the chameleon (\citealt{Khoury2003a,Khoury2003b,Mota2003}), where
the scalar field has a mass that depends on environment: the mass
is large in high density regions
(and therefore the scalar force is short range, or Yukawa suppressed), 
and small in low density ones.
The combined scalar $+$ tensor force is thus described by
an effective Newton's $G$ that is smaller in screened regions (such
as the solar system), and larger in unscreened regions such as voids.
As we will discuss below, the precise boundary between screened
and unscreened regions or objects is determined by the depth of
their gravitational potential.

Several recent papers discuss the observational implications
of the chameleon mechanism on structure formation 
(\citealt{Oyaizu2008,Schmidt2009}),
and on equivalence principle violations (Hui, Nicolis \& Stubbs 2009; 
\citealt{Hui2010}). 
The latter arises when one 
compares the motion
of a screened versus an unscreened object 
in an unscreened environment.
A screened object does not
respond to the scalar field, while
an unscreened object does. Therefore, they fall at different rates.
Screening affects not only the response of objects to the
scalar, but also the sourcing of the scalar field itself.
Namely, a screened object does not source the scalar field, while
an unscreened object does. This will be important for our discussion below.

In this paper, we show that stellar evolution is modified
in such scalar-tensor theories, and in particular the colors and 
luminosities of red giant branch (RGB) stars are measurably affected.  
Precision measurement of red giants in distant galaxies would place
strong constraints on theories that screen by the chameleon mechanism.  
The key point is that while a red giant's core is expected to be screened,
its envelope could well be unscreened depending on the parameters of
the theory. We focus on red giants both because this sort of effect
is most pronounced in them, and because they can be observed at 
great distances. But it should be mentioned that main sequence stars
in general will also be affected to some extent by such spatial variations
of the effective 
$G$. It is also worth mentioning that another screening mechanism,
known as the symmetron (\citealt{Hinterbichler2010}),
has a lot in common with the chameleon, namely that screened and unscreened
objects are distinguished by their gravitational potential. We expect
our conclusions on the modification of the red giant's structure to apply
to the symmetron case as well.
On the other hand,
theories that make use of the Vainshtein screening mechanism,
such as DGP (\citealt{DGP}) and the galileon (\citealt{Nicolis2009}),
are not expected to greatly alter red giants' structure, because
the screening extent is much less localized than the chameleon or 
the symmetron.

We begin by giving a brief
discussion of the chameleon mechanism in \S\ref{sec:scalar}, focusing
on how it operates inside a star, but also reviewing certain
standard results. 
We then use an appropriately modified version of the
stellar evolution code, MESA, \citep{Paxton+10} to calculate the effects of 
modified gravity on RGB stars in \S\ref{sec:structure}.  Finally, we
discuss how constraints on these modified gravity theories can be derived from
careful measurements of red giants in unscreened galaxies and
close in \S\ref{sec:discussion}. 

\section{The Scalar Field}\label{sec:scalar}

The scalar $\varphi$ in a chameleon theory is
described by
\begin{equation}\label{eq:scalar potential}
  \nabla^2 \varphi = \frac{\partial V}{\partial \varphi} + \alpha 8\pi G \rho,
\end{equation}
where $V$ is the self-interaction potential, $\alpha$ is the scalar
coupling, and $\rho$ is the mass density. Our notation and normalization
follows that of \cite{Hui+09}. For instance, $\alpha = 1/\sqrt{2}$ means
the (unscreened) 
scalar force between two masses has exactly the same strength
as that mediated by the graviton; $\alpha > 0$ in
our convention. We work in Einstein frame.

The potential $V$ is typically chosen such that it is large
for small $\varphi$ and small
for large $\varphi$, such as in equation (\ref{V}) below. 
As a result, the equilibrium value for the scalar field 
in a high density environment is small, and corresponds to a large mass,
which means it is Yukawa suppressed, i.e. screened.
Conversely, in a low density environment, the equilibrium value for $\varphi$
is large, and the scalar field has a small mass and is unscreened.
We will use the symbol $\varphi_*$ to denote the equilibrium value
at cosmic mean density $\rho_m$.

For definiteness, we adopt a potential $V$ of the form:
\begin{eqnarray}
\label{V}
V(\varphi) = B + {A \over \varphi^n},
\end{eqnarray}
where $B$, $A$ and $n$ are constants.\footnote{Our convention
is such that $V$, and therefore $B$ and $A$, have dimension of 
$1/{\,\rm time}^2$, and $\varphi$ is dimensionless.}  
Imposing the condition $\varphi = \varphi_*$ for $\rho = \rho_m$ implies:
\begin{eqnarray}
A = 3 H_0^2 \Omega_m^0 {\varphi_*^{n+1} \over n} \alpha, 
\end{eqnarray}
where $H_0$ is the Hubble constant today, and $\Omega_m^0 = 0.3$ is
the matter density today.  $B$ should take a value such that 
$V(\varphi_*)$ accounts for the vacuum energy today:
\begin{equation}
B =3 H_0^2 \left[1 - \Omega_m^0 (1 + {\alpha \varphi_* \over n}) \right].
\end{equation}
However, as we are interested in the derivative of $V$ with 
respect to $\varphi$, 
$B$ is irrelevant for our calculation.
As we will see, our conclusions are insensitive
to the precise form of the potential, for instance, the choice of $n$.


For an extended, spherical, object -- say a star -- described by some
mass density field $\rho(r)$, the scalar field profile is determined by
equation (\ref{eq:scalar potential}), which can be solved numerically as follows. Discretizing,
labeling radial position using the index $i$, and
using the potential outlined above, we can rewrite
equation (\ref{eq:scalar potential}) as
\begin{eqnarray}\label{eq:full dis equation}
\frac{\varphi_{i+1} + \varphi_{i-1}}{2} + \frac{\varphi_{i+1} - \varphi_{i-1}}{4} (\Delta\ln r) -  \frac {\eta}{2}\rho_i r_i^{2} \nonumber\\= \varphi_i - \frac{\eta}{2}\rho_m \left(\frac{\varphi_*}{\varphi_i}\right)^{n+1}r_i^{2},
\end{eqnarray}
where $\eta \equiv \alpha 8\pi G (\Delta\ln r)^2$.
Equation (\ref{eq:full dis equation}) with the appropriate boundary conditions ($d\varphi/dr = 0$ for small $r$ and $\varphi = \varphi_*$ for large $r$) is solved using a nonlinear Gauss-Seidel algorithm with a Newton-Raphson root find at every iteration \citep{Press+92}.
More concretely, given a guess for the scalar profile which
can be plugged into the left hand side of equation 
(\ref{eq:full dis equation}),
one can solve for $\varphi_i$ on the right.
Iteration then converges to the correct solution.

While this computation gives the most accurate answer
for a given $\rho(r)$, it is sufficiently 
costly that it cannot be used to 
construct hydrostatic models of stars, which involves
repeated calculations of $\rho(r)$ itself.
We therefore employ the following ansatz,
which is fairly accurate.
First, we divide up the star into three regions, a screened center, an unscreened 
"mantle", and an unscreened exterior. In the screened central region, 
the scalar field sits at the (small) local equilibrium value
at each radius:
\begin{equation}\label{eq:screened star}
\frac{d V}{d\varphi} \approx -\alpha 8\pi G \rho.
\end{equation}
Note that the screened center here might include
both the degenerate core of the red giant as well as
part of its envelope.
In the unscreened "mantle", the scalar field
starts to be driven by the density:
\begin{equation}\label{eq:unscreened star}
\nabla^2 \varphi \approx \alpha 8\pi G \rho.
\end{equation}
Finally, outside the star, the density is so much lower that
we can approximate
\begin{equation}\label{eq:outside star}
\nabla^2 \varphi \approx 0.
\end{equation}
The last approximation remains valid only for $r$ smaller 
than the Compton wavelength outside the star. 
In an unscreened environment, this Compton wavelength is much
larger than the size of the star, and we assume $\varphi$
asymptotes to $\varphi_*$ far away.
Essentially the same approximation scheme was worked out
in \cite{Khoury2003a}. The main difference from their solution
is that our $\rho(r)$ is not a top-hat, and
we have an extended region i.e. the mantle, 
where the scalar field is unscreened, but the density is 
non-negligible.

At the radial boundary, $r_{\rm scr}$, between the screened center and the unscreened mantle, 
let the field value and its derivative be
$\varphi_{\rm scr}$ and $k \varphi_{\rm scr}/r_{\rm scr}$,
where $k$ is a constant of order unity.
By demanding that $\varphi$ and its derivative are continuous at $r_{\rm scr}$,
we find 
the solution in the unscreened mantle (eq.[\ref{eq:unscreened star}]),
$r_{\rm scr} < r < R$ (stellar radius), to be:
\begin{equation}\label{eq:soln inside}
\varphi(r) = G\int_{r_{\rm scr}}^r dr' \frac{Q(r')}{r'^2} - k\varphi_{\rm scr}\frac{r_{\rm scr}}{r} + (1 + k)\varphi_{\rm scr},
\end{equation} 
where $Q(r) \equiv 8\alpha\pi\int_{r_{\rm scr}}^r \rho(r') r'^2 dr'
= 2\alpha [M(r) - M(r_{\rm scr})]$ can be
thought of as the scalar charge inside $r$. Here, $M(r)$ is the mass
interior to $r$.
Outside the star $r > R$, the solution to equation (\ref{eq:outside star}) is 
\begin{equation}\label{eq:soln outside}
\varphi = \varphi_*- \frac A r,
\end{equation}
where $A$ is a constant. Demanding continuity and differentiability
at stellar radius $R$ determines $A$ and $r_{\rm scr}$:
\begin{eqnarray}\label{eq:ansatz1}
&& A \approx G Q(R) \, ,\\
\label{eq:ansatz2}
&& \frac {GQ(R)}{R} + \int_{r_{\rm scr}}^{R} \frac {GQ(r)}{r^2} \approx \varphi_* \, ,
\end{eqnarray}
where we have ignored terms involving $\varphi_{\rm scr} \ll \varphi_*$.
The acceleration due to the scalar force is given by $- \alpha \Bnabla 
\varphi$.
The combined scalar $+$ gravitational radial acceleration $g_{\rm eff}$ is thus
\begin{eqnarray}
\label{eq:totalforce}
&& g_{\rm eff} = -\frac {G\left(M(r) + \alpha Q(r)\right)}{r^2} \quad {\rm for}\,
r > r_{\rm scr} \, ,\nonumber \\
&& g_{\rm eff} = -{GM(r) \over r^2} \quad {\rm for}\, r < r_{\rm scr} \, .
\end{eqnarray} 
The scalar force at $r > r_{\rm scr}$ is sourced only by the unscreened portion
of the star, i.e. the scalar charge receives contribution only from
the mantle. Within the screened region $r < r_{\rm scr}$, the scalar
force is Yukawa suppressed, and therefore 
the only operating force is gravitational
(in the Einstein frame sense).
The difference in how the total acceleration
relates to $M(r)$, for $r > r_{\rm scr}$
and $r < r_{\rm scr}$, can be interpreted as
a spatial variation of an {\it effective} $G$.
The $G$ used in our expressions throughout is a constant,
and corresponds to the value observed in the solar system.

Equations (\ref{eq:ansatz1}) and (\ref{eq:ansatz2}) can be rewritten 
as 
\begin{equation}
\label{eq:ansatz3}
\int^{\infty}_{r_{\rm scr}}\frac {GM(r)}{r^2} dr - \frac {GM(r_{\rm scr})}{r_{\rm scr}} \approx \frac {\varphi_*}{2\alpha}
\end{equation}
The first term is simply (minus) the value of 
the gravitational potential at $r = r_{\rm scr}$ (fixing
the potential to be zero at infinity).  
The second term is of a similar order. 
Equation (\ref{eq:ansatz3}) is a more accurate
version of the common statement that the gravitational
potential defines the boundary between screened and unscreened
regions -- regions with deeper gravitational potential
than $\varphi_*/2\alpha$ 
are screened
($-$ grav. pot. $\gsim \varphi_*/2\alpha$),
while regions with
shallower potential 
are not
($-$ grav. pot. $\lsim \varphi_*/2\alpha$). 
Note that only two parameters of the chameleon theory are
relevant for predicting the scalar force: $\alpha$ which
controls the strength, and $\varphi_*/2\alpha$
which controls where screening takes place. Details of the 
potential $V$ are unimportant.

\begin{figure}
  \plotone{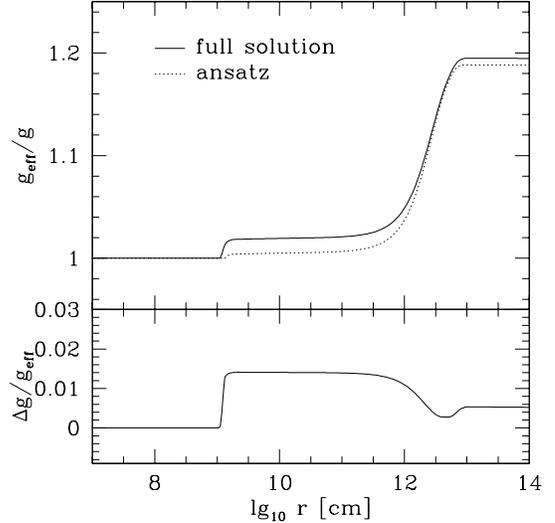}
  \caption{Comparison of the total force (gravitational and scalar)
    using a full numerical solution and our analytic ansatz, for 
    the density profile of a 1
    \Msun\ red-giant (Figure \ref{f:structure}), and 
    $\varphi_*/2\alpha = 10^{-6}$. The numerical solution
    uses a potential $V$ (eq. [\ref{V}]) with $n=2$. Other choices
    of $n$ gives very similar results.
    The upper panel shows the total force, $g_{\rm eff}$, normalized by 
    the gravitational (non-scalar) force, $g=GM(r)/r^2$. 
    The enhancement in the {\it effective} gravitational
    force (i.e. total) 
    is $\approx 20\%$ in the outer envelope due to the scalar
    field (a result of the choice $2\alpha^2 = 1/3$; the
    enhancement does not reach $1/3$ because of the screened
    core).
    The screening radius $r_{\rm scr}$ is located at
    about $10^9$ cm. 
    The lower panel shows the fractional
    difference in the total force between the numerical and analytic
    solutions. It is at the percent level.
}
  \label{f:gravity}
\end{figure}

As a check of the accuracy of our ansatz,
in Figure \ref{f:gravity}, we compare the total force
according to our ansatz (eq. [\ref{eq:totalforce}]) against
that from an exact numerical calculation (eq. [\ref{eq:full dis equation}]).
We use the same stellar model as in Figure
\ref{f:structure} (i.e. same $\rho(r)$).
Here, as in elsewhere in this paper unless
otherwise stated, we use $\alpha = 1/\sqrt{6}$ which is
the value for f(R) (\citealt{Carroll2004}).\footnote{Our treatment goes beyond f(R) theories and holds true for any scalar-tensor theory with a scalar interaction of the chameleon (or symmetron) kind.  However, it is useful to note the relation between $\varphi_*/2\alpha$ and the analogous parameter in a f(R) theory that arises from a modified action of the type $R\rightarrow R + f(R)$, where $R$ is the Ricci scalar. This relation is: $\varphi_* /(2\alpha) = - (df/dR)/(4\alpha^2) = -1.5 (df/dR)$.} Note however this value is
not protected by symmetry, the generic expectation
is $\alpha \approx O(1)$ (\citealt{Hui2010}).
Figure \ref{f:gravity} shows that our ansatz 
is accurate at the level of $\approx 1\%$.
We will use this ansatz to perform 
our stellar evolution calculation. 


\section{Red Giant Structure in Chameleon Gravity}\label{sec:structure}

\begin{figure}
  \plotone{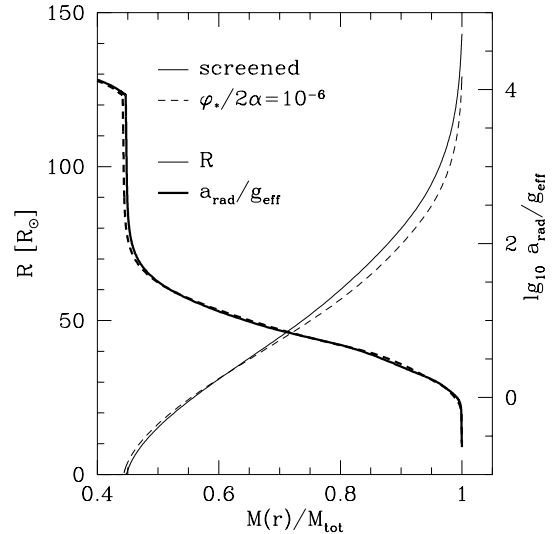}
  \caption{Radius (thin lines) as a function of mass fraction for a 1
    \Msun\ red giant for a star near the TRGB, with $\varphi_*/2\alpha =
    10^{-6}$ (dashed line; core screened) and 0 
    (solid line; entirely screened i.e. GR),
    and $2\alpha^2 = 1/3$.  Also plotted is the ratio between the radiative 
    acceleration, $a_{\rm rad}$, and effective gravity, $g_{\rm eff}$, in thick lines.  
    The age of the two stars is respectively
    $10.9$ and $11.9$ Gyrs for the partially and completely screened case
    (the age is chosen to yield the same luminosity of $2000 \, \Lsun$). 
    Note the $10\,\Rsun$ difference in photospheric
    radius between these two cases.}
  \label{f:structure}
\end{figure}

\begin{figure*}
  \plottwo{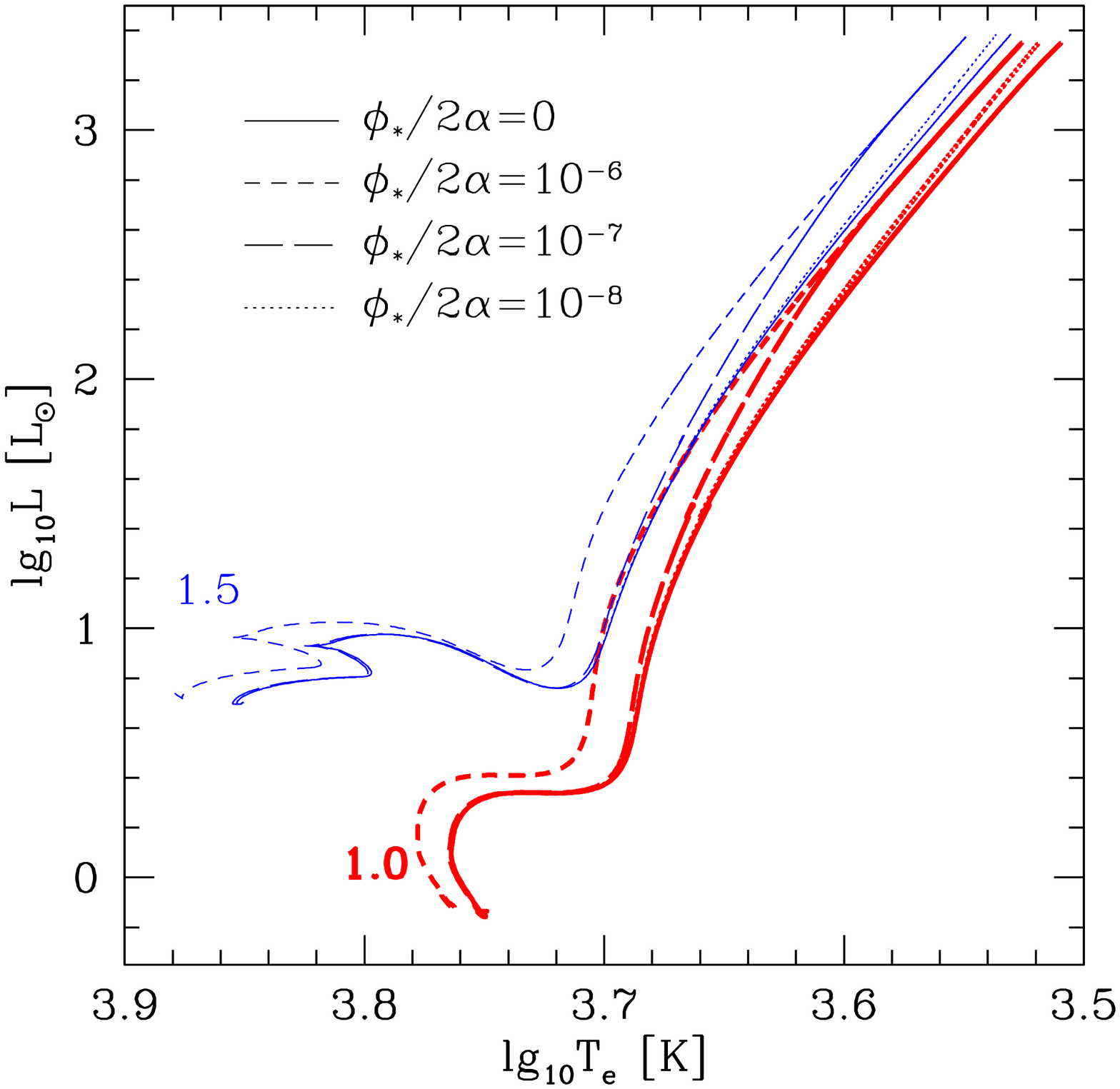}{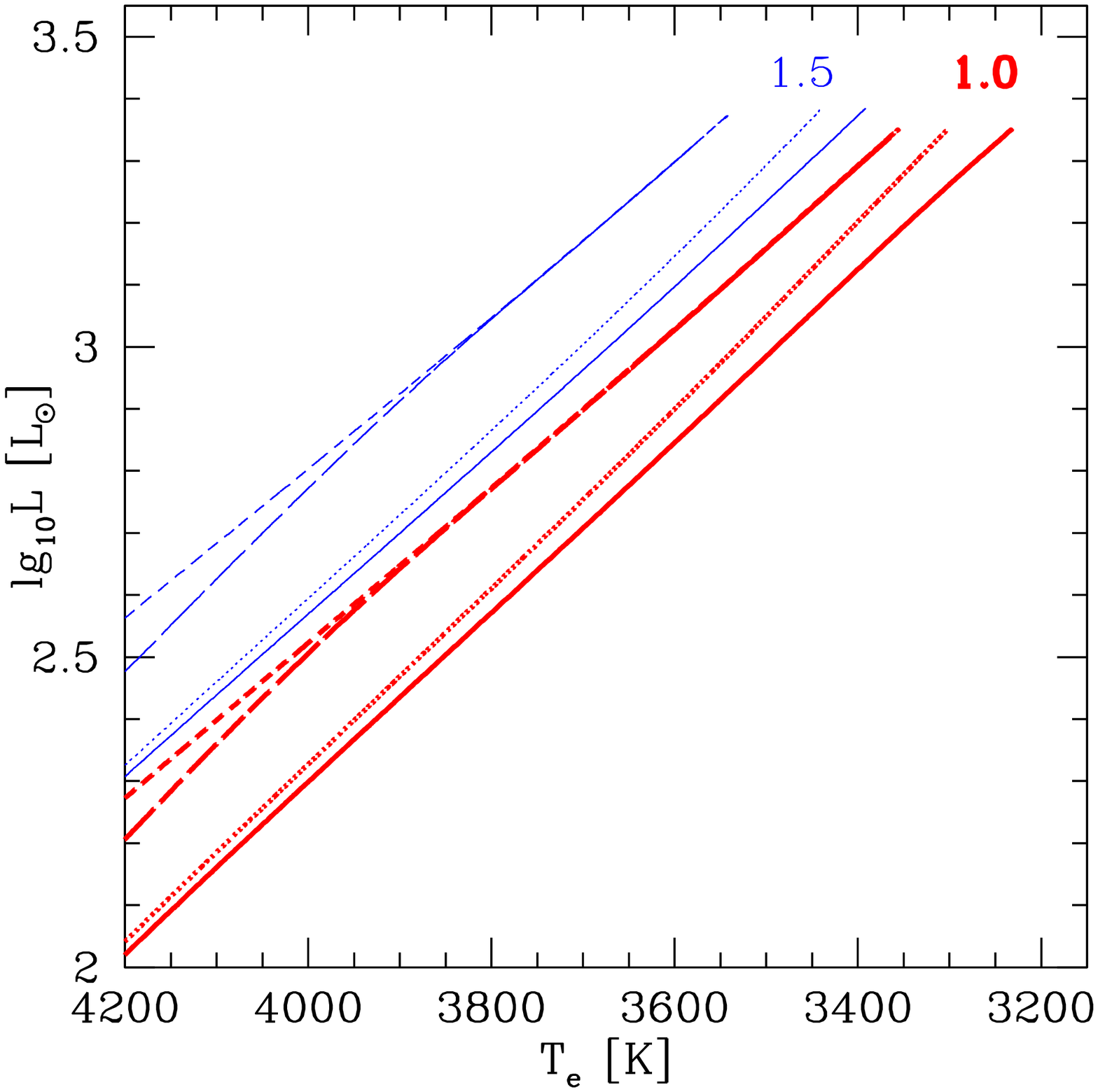}
    \caption{Left plot shows evolutionary track of stars of various masses (denoted in
    units of solar masses for each track: $1\Msun$ in thick red lines and $1.5 \Msun$ in thin blue lines) for different values of
    $\varphi_*/2\alpha = $ 0 (solid lines; scalar completely screened i.e. GR), 
    $10^{-6}$
    (short-dashed lines), $10^{-7}$ (long-dashed lines), and $10^{-8}$
    (dotted lines), all with $2\alpha^2 = 1/3$ appropriate for f(R) gravity. 
    Right plot is a closeup of the star on the RGB highlighting the difference 
    between the completely screened star, and the partially screened ones.  
    Note the difference in $T_e$ between the two is $\approx 130$ K.}
  \label{f:hr}
\end{figure*}

We modified the stellar evolution code MESA \citep{Paxton+10} to account for the scalar field changes to the local
gravity and calculate stellar evolution under these conditions.  While the scalar field modified many different aspects of stellar evolution, we focus on the RGB phase as these stars are sufficiently bright that they can be observed in distant galaxies (see our discussion in \S\ref{sec:discussion}). In Figure \ref{f:structure} we show the structure of a $\approx
2\times 10^3\,\Lsun$ RGB star that evolved from a 1 $M_{\odot}$ main sequence star with $\varphi_*/2\alpha = 10^{-6}$ (i.e. screened at its core)
and 0 (i.e. screened in its entirety, meaning GR limit).
Solar metallicity is assumed, as in the rest of the paper.
The difference in photospheric radius between 
the two stars is $\approx 10\,\Rsun$ with the completely
screened star being larger ($R\approx
145\,\Rsun$).  Because of the smaller radius of the
scalar-field influenced star, its effective temperature is 
larger by $\approx 150$ K ($T_{\rm eff} = 3395$ K vs $3258$ K).  


In Figure \ref{f:hr}, we plot the HR diagram for 1 (red) and 1.5 (blue) $\Msun$
stars with different degrees of screening.
We evolve a star from the
zero-age main sequence (ZAMS) up to the tip of the red giant branch
(TRGB) for $\varphi_*/2\alpha = $ 0 (solid lines; no scalar correction
i.e. complete screening),
$10^{-6}$ (short-dashed lines), $10^{-7}$ (long-dashed lines), and
$10^{-8}$ (dotted lines).  We have not considered values smaller than $\varphi_*/2\alpha < 10^{-8}$ as the potential depth ($v_{\rm cir}^2/c^2$) of dwarf galaxies are typically $\sim 10^{-8}$. This means
that cases where $\varphi_*/2\alpha < 10^{-8}$ are subject to blanket
screening: a host galaxy with a potential as shallow as that
of a typical dwarf is sufficient to screen the scalar for the star of 
interest
(see \S \ref{sec:discussion} for more discussions on blanket screening).

We note the following features.  For $\varphi_*/2\alpha = 10^{-6}$, 
significant departures from the main sequence is immediately apparent, 
whereas for $\varphi_*/2\alpha = 10^{-7}$ and $10^{-8}$, significant 
deviation do not appear until the RGB phase. 
That is to say, the scalar field affects the outer envelope for a 1 $\Msun$
 star even on the main sequence, effectively increasing the star's mass
(i.e. increasing its photospheric temperature for the same luminosity).   
This effect is reduced and harder to observe
for smaller values of $\varphi_*/2\alpha$.  
We do not comment on this effect here, choosing instead to focus on RGB stars, but it potentially allows for yet another probe of modified gravity, albeit for larger values of $\varphi_*/2\alpha$ ($\sim 10^{-6}$). 
Second, all three $\varphi_*/2\alpha$'s show measurable differences from the the fully screened case (i.e. GR limit) on the RGB.   Namely, the scalar-field influenced 1 \Msun\ case has nearly the same effective temperature as the 1.5\Msun\ RGB for $\varphi_*/2\alpha = 10^{-6}$ and $10^{-7}$ , i.e., the scalar-influenced star is hotter by $\approx 150$ K (this effect is smaller -- $\Delta T_{\rm eff} \approx 60$ K -- for the $\varphi_*/2\alpha = 10^{-8}$ case). 

To develop a better qualitative understanding of why red giants are such good probes of chameleon gravity, we plot the ratio of the radiative acceleration, $a_{\rm rad}$, and the effective gravity,  
$g_{\rm eff}$ (total scalar $+$ gravitational acceleration), as a function of mass fraction in Figure \ref{f:structure}. Here we define $a_{\rm rad} = \kappa L/4\pi r^2 c$ as a function of the {\it total} luminosity, $L$, where $\kappa$ is the Rosseland mean opacity.  As $a_{\rm rad} > g_{\rm eff}$ for the vast majority of the envelope, the energy flux is not carried outward by radiation, but rather by convection \citep{Kippenhahn+90,Paczynski69}.  Only near the photosphere, where $a_{\rm rad} = g_{\rm eff}$, can radiation carry away the star's luminosity.  It is this condition $a_{\rm rad} = g_{\rm eff}$ that determines the position of the photosphere for both the fulled screened and 
partially screened (scalar-influenced) star.  

The opacity in the envelope is dominated by H$^-$ opacity, which scales like $T^9$ for $3000 \lesssim T \lesssim 6000$ K \citep{Hansen+04}.  The scaling between effective gravity of the red giants (due to scalar fields) and the photospheric temperature of $\epsilon_{\rm eff} \sim 9\Delta T_{\rm ph}/T_{\rm ph} $ then gives 
\begin{equation}
  \Delta T_{\rm ph} \approx 110 \left(\frac {T_{\rm ph}}{4000\,{\rm K}}\right) \left(\frac {\epsilon_{\rm eff}}{0.25}\right)\ {\rm K},
\end{equation}
which roughly matches what we are finding in the full calculation. 

\begin{figure}
  \plotone{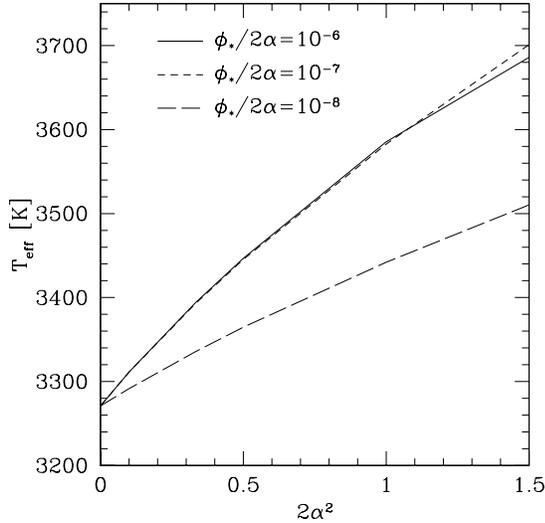}
  \caption{The effective temperature as a function of the scalar coupling, $2\alpha^2$, for a $L = 2000 \Lsun$, $1 \Msun$ RGB star.   For a red giant that is not under the influence of a scalar field, i.e., $2\alpha^2 = 0$, $T_{\rm eff} \approx 3270\,{\rm K}$.  As the coupling of the scalar field increases, $T_{\rm eff}$ increases, a result of the smaller size of the RGB star.}
  \label{f:teffalpha}
\end{figure}

Finally in Figure \ref{f:teffalpha}, we consider the effect of different values of the scalar coupling, $\alpha$, on RGB structure for three different values of $\varphi_*/2\alpha = 10^{-6},\ 10^{-7}$ and $10^{-8}$.  We plot the effective temperature of a $2000 \Lsun$ RGB star for different values of the scalar coupling.  The case of no scalar coupling $2\alpha^2 = 0$ is the
GR limit, where $T_{\rm eff} \approx 3270$ K.  As $\alpha$ increases, the effective temperature increases.  For instance the effective temperature is $\approx 3400$ K for $2\alpha^2 = 1/3$ (the f(R) gravity case we studied above), which is an increase of $\approx 130$ K over the unmodified case.  For $2\alpha^2 = 1.5$, $T_{\rm eff} \approx 3700$ K, an increase of over $400$ K above the fully screened case.

\section{Discussion}\label{sec:discussion}

As we have mentioned, we have mainly focused on RGB stars as they are observable in distant galaxies due to their large luminosity.  Indeed tip of the RGB stars are seen out to $\approx 30$ Mpc \citep{Mager+08} and used for accurate distance measurements (see the review \citealt{Freedman+10}).  We have shown above that their temperature is also very susceptible to the effects of scalar fields.  We now discuss their use as constraints on modified gravity theories. 

First let us discuss the precision with which their temperature can be measured.
\cite{Ferraro+06} (see also \citealt{Cassisi07}) collected high quality J, H, and K photometry 
of 28 Galactic globular clusters to measure the effective temperatures of RGB stars at fixed bolometric 
luminosities.  Their measurement uncertainty of $\approx \pm 30-60$ K suggests that measuring 
an effective temperature difference of $\approx 150$ K is quite viable.  
For a scalar-matter coupling $\alpha$ larger than $1/\sqrt{6}$, 
the expected temperature difference should be even larger.

We caution other effects are also important in influencing the temperature of RGB stars.
For instance, metallicity can change the effective temperature of a RGB star by 
a few hundred K (see Figure 4 of \citealt{Cassisi07} or Figure 2 and 3 of \citealt{Ferraro+06}).
Fortunately, metallicity effects are well modeled using modern stellar evolution codes 
(again see Figure 4 of \citealt{Cassisi07}).  A constraint on the metallicity of these stars, e.g., spectroscopic measurements, will help mitigate these effects. 
There are also additional effects that might contribute, i.e., uncertainties in mixing-length theory, 
conductive opacities, etc, but these are being mitigated via comparison to local RGB stars.

The effects of modified gravity on stellar evolution (for $\phi_*/2\alpha \sim 10^{-8} - 10^{-7} $) would also manifest itself in deviation of the RGB branch under modified gravity (see Figure \ref{f:hr}). This deviation would be a unique signature of modified gravity compared to metallicity variations.  Constraining this deviation in ``clean'' system such as globular clusters around unscreened galaxies would also place limits on the parameter of the chameleon mechanism.  

As is apparent from Figure \ref{f:hr}, the mass of the initial star may also shift the color of these TRGB stars so that they appear bluer.  Namely the degeneracy between scalar field effects and mass of the RGB star needs to be broken.  An examination of Figure \ref{f:hr} suggests several methods by which this can be done.  If $\varphi_*/2\alpha$ is small, the deviation from the RGB track due to the effect of the scalar field can be used to directly test for the strength of the scalar field.   However, for a sufficiently large $\varphi_*/2\alpha$, this happens sufficiently early in the RGB phase that the deviation might be missed.  In that case, the mass of these RGB stars would need to be established and this can happen in several ways.  For instance, if the stellar population is relatively uniform and a clean turnoff mass can be identified, this would 
put a strong constraint on the mass of the progenitor.  
Note that the RGB track for the scalar influenced 1 $\Msun$ traces the (fully) screened 1.5  
$\Msun$ track fairly closely.  
However, the positions of the turnoff for 1 and 1.5 $\Msun$ stars are significantly 
different.  


We now discuss the current constraints on chameleon theories,
and examine what kind of host galaxies are needed to improve them.
Stringent tests of GR in the solar system tell us that the
scalar must be screened within it. 
The simplest interpretation is that the solar system
is screened by virtue of its residing inside the Milky Way (MW),
which has a self gravitational potential of $\sim - 10^{-6}$.
\footnote{By coincidence, the sun has a 
similar gravitational potential, $\sim - 2 \times 10^{-6}$.
} This suggests $\varphi_*/2\alpha \lsim 10^{-6}$ so that the MW,
and therefore its constituent solar system, is screened.
\footnote{Another option is of course that the scalar-matter coupling
$\alpha$ is very close to zero. This is the uninteresting limit of
the scalar force being practically invisible to everything else.
Our discussion assumes instead gravitational strength
coupling of $\alpha \sim O(1)$, including for instance the f(R)
value of $\alpha = 1/\sqrt{6}$.
Observational constraints on chameleon theories ultimately
translate into limits on the plane of
$\varphi_*/2\alpha$ and $\alpha$
(see Fig. 3 of \citealt{Hui+09}).
}
This could possibly be evaded by saying that
the MW is screened not so much by its self gravitational potential,
but by virtue of its residing in the local group.
But data and constrained realizations suggest the local group
has a gravitational potential rather similar to that of the MW
itself (\citealt{Klypin2003}). Neighbors of the local group are
unlikely to change the picture significantly -- for instance,
both the Virgo cluster and the local void contributes a potential
at the MW of the order of $10^{-6}$, with opposite signs 
(\citealt{Peebles2010}).
Structure formation offers a constraint independent of
such MW considerations: $\varphi_*/2\alpha \lsim 10^{-4}$
from the observed cluster abundance (\citealt{Schmidt2009b}).
We will frame our discussion mainly
in terms of improving the MW constraint of $10^{-6}$, though
obviously the improvement will be even greater when compared
against the more conservative structure formation limit.\footnote{
Indeed, if $\varphi_*/2\alpha$ were larger than about
$4 \times 10^{-5}$, even the red giant core would start to
be unscreened (assuming its host galaxy is unscreened), and
the predicted modifications to the luminosity-temperature relationship
would be greater. A discussion of the behavior of stars under strongly modified physical constants (including $G$) is found in \cite{Adams08}.}

Red giants located in a host galaxy with a gravitational potential
shallower than the MW can improve upon the existing constraint.
For instance, a $100$ km/s host galaxy could
push the limit on $\varphi_*/2\alpha$ to $10^{-7}$;
a $30$ km/s dwarf could push it to $10^{-8}$.
The important point is to avoid blanket screening, i.e.
avoid host galaxies that are screened by their surroundings,
such as by virtue of being situated inside a cluster or massive
group.\footnote{In other words, screening communicates
from host to members. A red giant inside a galaxy inside
a cluster is screened because of the deep potential of the cluster,
even if the galaxy does not have a deep {\it self}-potential.}
This means that for red giants located within a $100$ km/s galaxy
to be useful, we would like to make sure the galaxy's surroundings
contribute a potential no deeper than $- 10^{-7}$. 
This should be achievable if the galaxy is located at a distance
of more than a few Mpc from the MW\footnote{Our calculations of the screening effect in an NFW profile \citep{Navarro+96} suggest that the blanket screening of the MW halo is important up to a few times $r_{200}$, so that the local dwarfs are likely not useful for this measurement.}, avoiding the local sheet and into
the general direction of the local void (\citealt{Klypin2003}).
For red giants located within a $30$ km/s dwarf to be useful,
we should make sure the galaxy's neighbors contribute
a potential no deeper than $-10^{-8}$. For this, one could
venture deeper into the local void (\citealt{Peebles2010})
, or go to the field or voids 
that are further away from nearby concentrations of galaxies
($\gsim 10$ Mpc,   
\citealt{Szomoru+96a,Szomoru+96b,Rojas+04,vandeWeygaert2009,Stanonik2009}).
Obviously, an accurate mapping of the gravitational potential
of the local universe ($\lsim$ tens of Mpc) would be highly
desirable in determining which galaxies are likely to be
unscreened by environment.

In this paper, we have shown that RGB stars in galaxies with
shallower potential than the MW can be used to improve constraints
on modified gravity theories that invoke the chameleon or symmetron
mechanism.
We have shown that RGB stars with an unscreened envelope (but
generally screened core) are more compact, and hence hotter (by $\approx 150$ K) than completely screened RGB stars (like the ones in the MW) at the same 
luminosity.  
This temperature difference should be measurable in distant 
unscreened galaxies.
Dwarf galaxies in the field or in voids will give us the strongest
limits, improving the current solar system or MW constraint by
2 orders of magnitude, and current structure formation constraints
by 4 orders of magnitude.

\acknowledgements
We thank Bill Paxton for helping us set up MESA and his
patience with answering our many questions regarding its design.  Without his help, this work would not have
been possible. We also thank B. Madore for useful discussions. P.C. is supported by the Canadian Institute for
Theoretical Astrophysics. L.H. is supported by the DOE (DE-FG02-92-ER40699) and NASA
(09-ATP09-0049), and thanks Hong Kong University,
New York University and
the Institute for Advanced Study for hospitality.
This research has made use of NASA's Astrophysics Data System.
\\

\bibliographystyle{apj} 
\bibliography{ms}

\end{document}